\documentclass[aps,pre,reprint,amsmath,amssymb, superscriptaddress,showpacs]{revtex4-1}
\usepackage{graphicx}

\begin{document}


\title{Brownian motion of a self-propelled particle}


\author{Borge ten Hagen}
\email[]{bhagen@thphy.uni-duesseldorf.de}
\affiliation{Institut f\"ur Theoretische Physik II: Weiche Materie, Heinrich-Heine-Universit\"at D\"usseldorf, Universit\"atsstr.~1, 40225 D\"usseldorf, Germany}
\author{Sven van Teeffelen}
\affiliation{Department of Molecular Biology, Princeton University, Princeton, NJ 08544, USA}
\author{Hartmut L\"owen}
\email[]{hlowen@thphy.uni-duesseldorf.de}
\affiliation{Institut f\"ur Theoretische Physik II: Weiche Materie, Heinrich-Heine-Universit\"at D\"usseldorf, Universit\"atsstr.~1, 40225 D\"usseldorf, Germany}


\date{\today}

\begin{abstract}
Overdamped Brownian motion of a self-propelled particle is studied  by solving the
Langevin equation analytically. On top of translational and rotational diffusion, in the context of the presented model,  
the ``active'' particle is driven  along its 
internal orientation axis. 
We calculate  the first
four moments of the probability distribution function for displacements
 as a function of time for a spherical particle with isotropic translational diffusion
as well as for an anisotropic ellipsoidal particle. In both cases the translational and rotational motion is either unconfined or confined to one or two dimensions.
A significant non-Gaussian behavior at finite times $t$ 
is signalled by a non-vanishing kurtosis $\gamma (t)$.
To delimit the super-diffusive regime, which occurs at intermediate times, two time scales are identified. For certain model situations a characteristic $t^3$~behavior of the mean square displacement is observed. 
Comparing the dynamics  of real and artificial microswimmers like bacteria or
catalytically driven Janus particles to our analytical expressions reveals
whether their motion is Brownian or not. 
\end{abstract}


\maketitle


\section{\label{Einleitung}Introduction}
There are numerous realizations of 
self-propelled particles~\cite{Ramaswamy,Lauga_review:09} in nature 
ranging from
bacteria~\cite{Berg:72,Berg:90,DiLuzio:05,Lauga:06,Hill:07,Shenoy:07,Tailleur:09,Leptos:09} and
spermatozoa~\cite{Riedel:05,Woolley:03,Friedrich} to artificial colloidal
microswimmers. The latter are either catalytically 
driven~\cite{Dhar:06,Walther:08,Baraban:08,Baraban2,Golestanian:09,Popescu:09} or 
navigated by external magnetic fields~\cite{Dreyfus:05,Ghosh:09,Belkin:09},
but also biomimetic propulsion mechanisms can be exploited~\cite{Fery:08}.
On the macroscopic scale, vibrated polar granular
rods~\cite{Kudrolli:07,Daniels:09,Aronson:01} and even
pedestrians~\cite{Obata:05} provide more examples of ``active'' particles~\cite{Loreto:09,Dunkel:09}.
A suitable framework for theoretical modeling of self-propellers is provided by
the traditional Langevin theory of an anisotropic particle with
translational and orientational  diffusion including an effective
 internal force~\cite{footnote_forcefree,Lobaskin:08} in the overdamped Brownian dynamics \cite{Teeffelen_PRE,footnote_Erdmann}. 
The direction of the theoretically assumed internal propulsion force (corresponding to an imposed mean propagation speed)  
fluctuates according to rotational
Brownian motion~\cite{Doi_Edwards_book,HL_Cyl,Klein}. 
It is a challenging question whether real self-propelled particles
can at least in a rough way be covered according to this simple Brownian picture. While for ``passive''
ellipsoidal particles  a comparison  revealed  very good agreement with the 
picture of Brownian dynamics \cite{Han:06,Han:09,Nagele:96}, this has never been undertaken for
self-propellers.
Any deviations
point to the relevance of hydrodynamic interactions, 
non-Gaussian noise, or fluctuating internal forces which are beyond simple
Brownian motion.

Despite its simplicity, the Brownian motion of anisotropic
particles \cite{Einstein,Perrin1,Perrin2} has only been considered in the absence of internal driving forces 
either in the bulk \cite{Han:06,Ribrault:07} or 
in an external force field derivable from a potential \cite{Grima:07}. For ``passive'' rodlike particles the  Smoluchowski-Perrin equation~\cite{Doi_Edwards_book, Dhont_book} has been solved exactly in two~\cite{Munk_epl}  as well as in three~\cite{Aragon:85} dimensions. 
With regard to self-propellers, so far analytical results are only available if the orientation vector is confined to two dimensions. For rodlike particles~\cite{Elgeti:09,Tao:05} the first two~\cite{Teeffelen_PRE}, and for spherical  particles the first 
 four \cite{Cond_Matt} moments of the
probability distribution function for displacements were calculated. In this paper, we close the remaining  gaps by presenting a comprehensive model and calculating the first four moments
of the displacement distribution function for all relevant situations. First, we provide analytical results for an anisotropic self-propelled Brownian particle in two dimensions.
 Furthermore, both the situations of 
an isotropic and of an anisotropic particle are extended to
the full three-dimensional case where the orientation vector
is unconfined.

Studying the mean square displacement reveals a super-diffusive regime at intermediate times, which is characterized by a $t^2$~time dependence for most cases and beyond that by a $t^3$~behavior for some special cases. Moreover, two  time scales that delimit the super-diffusive regime are identified. 
These can be extracted from the results for the mean square displacement or from the normalized fourth cumulant (kurtosis) $\gamma (t)$ of the probability distribution function for displacements, which measures the non-Gaussian behavior as a function of time $t$.
For small and very large times, the kurtosis vanishes indicating Gaussian behavior,
but due to both  particle
anisotropy and self-propulsion, $\gamma (t)$ is non-vanishing for intermediate times. 
While Han and coworkers~\cite{Han:06} found the kurtosis of ``passive'' particles to be positive for $t>0$ with a simple 
maximum at finite time~\cite{Han:06},
here we find that
a propulsive force tends to make the kurtosis negative. There is a rich structure
in $\gamma (t)$ revealing different non-Gaussian behavior at different time scales.
Quite generally, the propelling force induces a negative massive long-time tail in $\gamma (t)$
which tends to zero as $1/t$. This prediction can in principle be verified in experiments on self-propelled
particles.

The paper is organized as follows: In Sec.~\ref{Modell} we present and motivate the  various model situations that are considered in  Secs.~\ref{einwinkel} to \ref{zweiwinkele} of this paper. In each case the first four displacement moments are calculated analytically and the results are analysed based on appropriate figures.  Finally, we conclude and give an outlook on further expansion of our model in Sec.~\ref{Schluss}.

\section{\label{Modell}Remarks about the various model situations}
\begin{figure}
\includegraphics[width=0.45\textwidth]{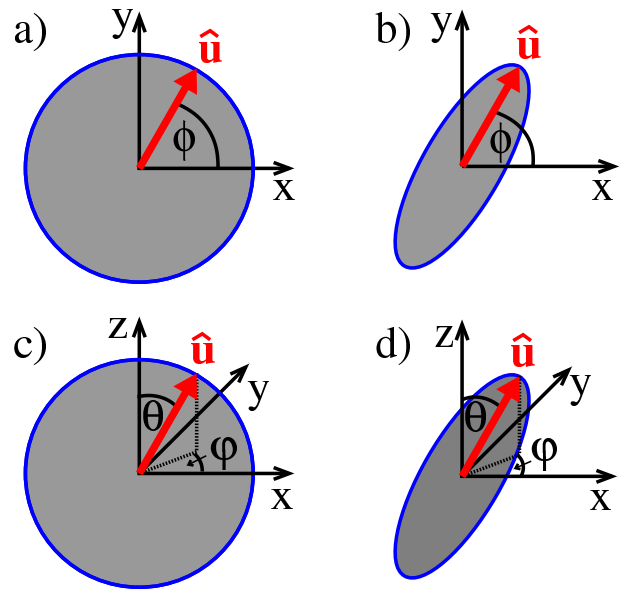}%
\caption{\label{fig:Modelle}(Color online) Sketch of the various  situations to which the model is applied: a) $(D,1,s)$-model, b) $(D,1,e)$-model, c) $(D,2,s)$-model and d) $(D,2,e)$-model. The notation is explained in the text. For $D=1$ in subfigures~(a) and (b) and for $D<3$ in subfigures~(c) and (d), the effective driving force along the particle orientation $\mathbf {\hat{ u}}$ is projected  onto the respective number of translational dimensions. To cover the $(1,1,s)$-model, for example, in subfigure (a) only the motion in the  $x$-direction is considered. }
\end{figure}

In this section, we  give an overview of the  situations to which the model is applied in this paper (see also Fig.~\ref{fig:Modelle}). In general, the model consists of an isotropic or anisotropic self-propelled  particle which undergoes completely overdamped Brownian motion. To describe the propulsion mechanism on average, we theoretically assume an effective internal force $\mathbf F=F\mathbf {\hat{ u}}$ that is included in the Langevin equation. The orientation vector $\mathbf {\hat{ u}}$ is introduced to specify the direction of the self-propulsion.  Depending on the number of translational degrees of freedom, in some of the cases to be covered this force is projected either onto a linear channel or onto a two-dimensional plane. To characterize the different situations depending on the number of degrees of freedom of the particle, we introduce the following notation: The $(D,d,\sigma)$-model refers to the situation with $D$ translational degrees of freedom and $d$ orientational degrees of freedom. The possible values for these parameters are $D\in\{1,2,3\}$ and $d\in\{1,2\}$. The parameter $\sigma\in\{s,e\}$ refers to the shape of the particle. While $\sigma = s$ relates to a spherical particle, for an ellipsoidal particle $\sigma = e$ is used.  When no specific value is given for one of these parameters, we refer to the group of models with an arbitrary value for  that parameter. 

We will first refer to the $(D,1,s)$-model, which is depicted in Fig.~\ref{fig:Modelle}(a) (theoretical investigation in Sec.~\ref{einwinkel}). This system consists of a self-propelled spherical particle whose rotational motion is constrained to a two-dimensional plane.  To study the behavior of the particle, we first refer to the one-dimensional translation in $x$-direction (Secs.~\ref{11s} to \ref{11ssquare}). In experiment, this situation can be achieved by confining swimmers by means of external optical fields~\cite{Lutz:04,Hanes:09}, for example.  After investigating this $(1,1,s)$-model the results can easily be transferred to the two-dimensional case ($(2,1,s)$-model), which is done in Sec.~\ref{21s}. This model situation is especially useful to describe the motion of a self-propelled particle on a substrate.
 
As the assumption of spherical particles is not justifiable in many experimental situations, the model is generalized to ellipsoidal particles (see Fig.~\ref{fig:Modelle}(b)) in Sec.~\ref{einwinkele} by investigating the $(D,1,e)$-model. The more complicated coupling between rotational and translational motion as opposed to spherical particles leads to qualitatively different results. 

Besides particles moving on a substrate with one orientational degree of freedom, we study freely rotating self-propelled particles. Here, the case of free translational motion in the bulk ($(3,2,\sigma)$-model) is interesting as well as situations in which the translation of the particle is constrained either to a linear channel ($(1,2,\sigma)$-model) or to a two-dimensional plane ($(2,2,\sigma)$-model). 
Again, a spherical particle (see Fig.~\ref{fig:Modelle}(c)) is discussed first (Sec.~\ref{zweiwinkel}), before the most general case of a freely rotating ellipsoidal particle (see Fig.~\ref{fig:Modelle}(d)) is considered in Sec.~\ref{zweiwinkele}.

\section{\label{einwinkel}Spherical particle with one orientational degree of freedom}
This section contains the theoretical considerations concerning the situation in Fig.~\ref{fig:Modelle}(a) ($(D,1,s)$-model). Although a spherical object is regarded here, one has to consider that a certain direction is specified through the theoretically assumed driving force $\mathbf F=F\mathbf {\hat{ u}}$. The two-dimensional motion of the self-propelled particle can be described by the coordinates $x$ and $y$  of the center of mass position vector $\mathbf{r}(t)= (x,y)$  and the angle $\phi$ between $\mathbf {\hat{ e}_x}$ and $\mathbf {\hat{ u}}=(\cos\phi,\sin\phi)$. 
Thus, the basic Langevin equations are given by  
\begin{align}
\label{Langevinx1}
\frac{\mathrm{d}\mathbf{r}}{\mathrm{d}t} &= \beta D_t \left[F \mathbf {\hat{ u}} - \mathbf{\nabla} U + \mathbf{f}\right]\,,\\
\label{Langevinphi1}
\frac{\mathrm{d}\phi}{\mathrm{d}t}&= \beta D_r \mathbf{g}\cdot \mathbf {\hat{ e}_z}\,.
\end{align}
Here, $\mathbf{f}(t)$ and $\mathbf{g}(t)$ are the Gaussian white noise random force and torque, respectively. They are characterized by $\langle f_i(t)\rangle =0$, $\langle f_i(t)
f_j(t')\rangle =2\delta_{ij}\delta(t-t')/(\beta^2 D_t)$, $\langle g_i(t)\rangle =0$, and $\langle g_i(t)g_j(t')\rangle =2\delta_{ij}\delta(t-t')/(\beta^2 D_r)$, where the indices $i$ and $j$ refer to the respective components, $\delta_{ij}$ is the Kronecker delta, and $\langle \ldots \rangle$ denotes a noise average. $U(\mathbf{r})$ is an external potential. The prefactors in Eqs.~(\ref{Langevinx1}) and (\ref{Langevinphi1}) consist of   the inverse effective  thermal energy $\beta=\left(k_\mathrm{B}T\right)^{-1}$ on the one hand and the  translational and
rotational short-time diffusion constants $D_t$ and $D_r$ on the other. In the case of a spherical particle with radius  $R$ these two quantities fulfill the relation $D_t/D_r = 4 R^2/3$, which is used in the following analytical expressions for the displacement moments. The Langevin equation~(\ref{Langevinphi1}) can easily be derived from the more general vector equation $(\mathrm{d}\mathbf {\hat{ u}})/(\mathrm{d}t)=\beta D_r \mathbf{g}(t) \times \mathbf {\hat{ u}}$. 

 As $\phi$ is a linear combination of Gaussian variables according to Eq.~(\ref{Langevinphi1}), the respective probability distribution function has to be Gaussian as well and proves to be 
\begin{equation}
\label{Verteilungphikonkret}
P(\phi ,t)=\frac{1}{\sqrt{4\pi D_\mathrm{r}t}}\exp\left({-\frac{(\phi-\phi_0)^2}{4D_\mathrm{r}t}}\right)\,,
\end{equation} 
where $\phi_0 \equiv \phi(t=0)$ is the initial angle. 
For the theoretical analysis, the two-dimensional motion in the  $xy$-plane can be split up into its components in $x$- and $y$-direction, respectively. In the following subsections, we first refer to the $x$-component of Eq.~(\ref{Langevinx1}).  As some calculations for  the $(1,1,s)$-model  have already been presented in~\cite{Cond_Matt} in more detail, we only summarize the most important results and briefly refer to systems with additional linear or quadratic potentials after that.

\subsection{\label{11s}The $(1,1,s)$-model}
Integrating the averaged Eq.~(\ref{Langevinx1}) for $U=0$ over time and considering only the $x$-component yields 
\begin{equation}
\label{Moment11}
\langle x(t)-x_0 \rangle =\frac{4}{3}\beta FR^2 \cos(\phi_0)\left[1-e^{-D_rt}\right]
\end{equation}
and 
\begin{align}
\label{Verschiebungsquadrat1}
\left\langle (x(t)-x_0)^2\right\rangle & = \frac{8}{3}R^2 D_r t + \left(\frac{4}{3} \beta FR^2\right)^2  \notag \\ 
& \quad \times \biggl[D_rt-1  +e^{-D_rt}   +\frac{1}{12} \cos(2\phi_0)  \notag \\
& \qquad \times \left(3-4e^{-D_rt}+e^{-4D_rt}\right)\biggr]
\end{align}
for the mean position and the mean square displacement. 

As usual the skewness $S$ and the kurtosis $\gamma$ are defined as  
\begin{equation}
\label{Schiefedef}
S=\frac{\left\langle(x-\langle x\rangle)^3\right\rangle}{\left\langle(x-\langle x\rangle)^2\right\rangle^{3/2}}
\end{equation}
and
\begin{equation}
\label{Woelbungdef}
\gamma=\frac{\left\langle(x-\langle x\rangle)^4\right\rangle}{\left\langle(x-\langle x\rangle)^2\right\rangle^2}-3\,,
\end{equation}
respectively.
A non-Gaussian behavior is manifested in  non-zero values for $S$ and $\gamma$. Using the notation $F_s^*=\beta RF$ for spherical particles  and a scaled time 
$\tau =D_rt$, the third and fourth moments are given by the analytical results
\begin{widetext}
\begin{align}
\left\langle \frac{(x(t)-x_0)^3}{R^3}\right\rangle  =  {\frac {32}{3}}\,F_s^*\tau\cos \left( \phi_0 \right)  \left( 1-{e^{-\tau}} \right)   +{\frac {64}{27}}\,{F_s^*}^{3} \biggl[& \cos \left( \phi_0 \right)\Bigl(3 \tau -{\frac {45}{8}}+\frac{5}{2} \tau{e^{-\tau}}
+{\frac {17}{3}} {e^{-\tau}} -\frac{1}{24} {e^{-4\,\tau}} \Bigr) \notag \\
& + \cos \left( 3\phi_0 \right) \Bigl( \frac{1}{24} - \frac{1}{16}{e^{-\tau}} +\frac{1}{40}{e^{-4\,\tau}}
-  {\frac {1}{240}}{e^{-9\,\tau}} \Bigr) \biggr]
\end{align}
and 
\begin{align}
\left\langle \frac{(x(t)-x_0)^4}{R^4}\right\rangle & =  {\frac {64}{3}}\,{\tau}^{2}+{\frac {256}{9}}\,{F_s^*}^{2}\tau \left[ {e^{-\tau}}+\tau-
1+\frac{1}{12}\,\cos \left( 2\phi_0 \right)  \left( {e^{-4\,\tau}}-4\,{e^{-\tau}}+3
 \right)  \right] \notag \\ 
 & \quad +{\frac {256}{81}}\,{F_s^*}^{4}\biggl[3{\tau}^{2} -
{\frac {45}{4}}\tau +{\frac {261}{16}} -5\tau{e^{-\tau}} -{\frac {49}{3}}\,{e^{-\tau}
} +\frac{1}{48}{e^{-4\,\tau}} \notag \\
& \qquad \qquad \qquad + \cos \left( 2\phi_0 \right) \biggl( \frac{3}{2}\tau\ -{\frac {19}{6}}+\frac{5}{3}\tau {e^{-\tau}} +{\frac {229}{72}
} {e^{-\tau}} -\frac{1}{30}\tau{e^{-4\tau}}  -{\frac {7}{450}}{e^{-4\tau}} +{\frac {1}{600}}{e^{-9\tau}} \biggr) \notag \\
& \qquad \qquad \qquad  + \cos \left( 4\phi_0 \right) \left({\frac {1}{192}}
 -{\frac {1}{
120}} {e^{-\tau}} +{
\frac {1}{240}} {e^{-4\,\tau}} -{\frac {1}{840}} {e^{-9\tau}} +{\frac {1}{6720}}{e^{-16\tau}}\right)\biggr]\,.
\end{align}
\end{widetext}

\begin{figure}
\includegraphics[width=0.45\textwidth]{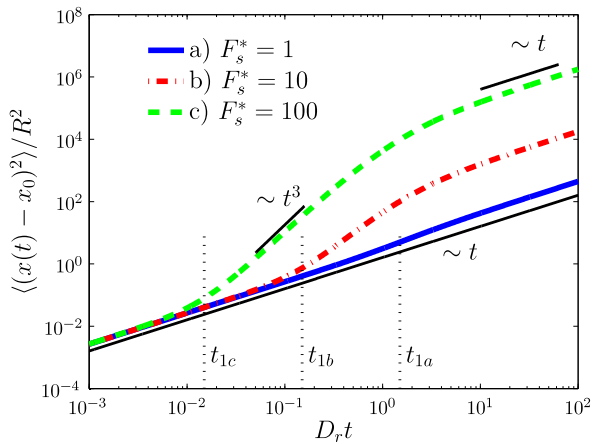}%
\caption{\label{fig:Moment2model11s}(Color online) Mean square displacement of a spherical particle with initial orientation angle $\phi_0 = 0.5 \pi$ for different values of $F_s^* = \beta R F$.  The different time dependences in the various regimes are illustrated by straight lines.}
\end{figure}
\begin{figure}
\includegraphics[width=0.45\textwidth]{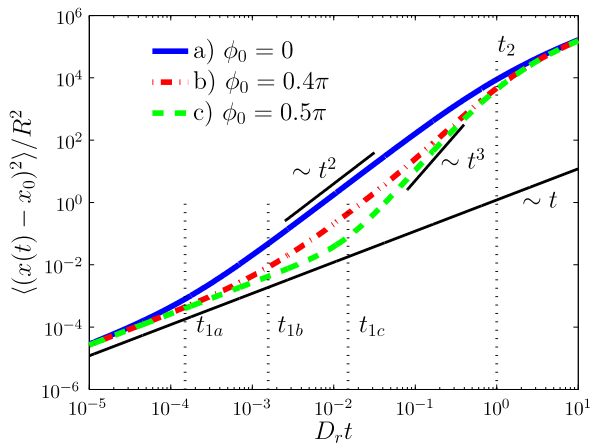}%
\caption{\label{fig:Moment2model11szeitskalen}(Color online) Mean square displacement of a spherical particle with effective force $F_s^*=100$ for different values of $\phi_0$. The regime of super-diffusive motion is defined by the time scales $t_1$ and $t_2$. While $t_2$ is independent of the initial angle $\phi_0$, the time scale $t_1$ is the bigger the more the initial orientation of the particle deviates from the $x$-direction.}
\end{figure}
In the case of large forces $\beta RF \gg1$, the particle motion (see Figs.~\ref{fig:Moment2model11s} and \ref{fig:Moment2model11szeitskalen}) is separated
into three qualitatively different time regimes, two diffusive regimes at short
and at large times, and a super-diffusive regime at intermediate times, which is characterized by a $t^2$ or a
$t^3$~behavior of the mean square displacement, depending on the
initial particle orientation.
These regimes are separated by the two time scales $t_1$ and $t_2$,
respectively: At early times $t\ll t_1$, the particle undergoes simple
translational Brownian motion, which is governed by the short time
translational diffusion term $(8/3) R^2 D_r t$ in Eq.~(\ref{Verschiebungsquadrat1}).  As seen in Figs.~\ref{fig:Moment2model11s} and \ref{fig:Moment2model11szeitskalen}  the
mean square displacement displays a crossover to an intermediate
super-diffusive regime at a time scale $t_1$, which, in turn, depends on the initial
orientation $\phi_0$ and the effective force $F_s^* = \beta RF$. In particular,
\begin{equation}
\label{ampl-eq}
t_1=\left\{\begin{array}{ll}
(3/2) (\beta R F D_r)^{-1}\,,&|\cos(\phi_0)|<1/\sqrt{\beta R F}\\
(3/2) [\cos(\phi_0) \beta R F]^{-2}D_r^{-1},&|\cos(\phi_0)|>1/\sqrt{\beta R F}\,.
\end{array}\right.
\end{equation}
If the initial orientation has a sufficiently large component parallel to
the $x$-axis, i.e., if $|\cos(\phi_0)|>1/\sqrt{\beta R F}$, the mean
square displacement displays a
crossover to a ballistic regime, which is governed by a scaling relation
$\langle(x(t)-x_0)^2\rangle\propto t^2$. The crossover is observed the
earlier the larger the inital force component $|\cos(\phi_0)| \beta R F$
parallel to the $x$-axis. On the contrary, if the initial particle
orientation points along or almost along the $y$-axis at $t=0$, i.e.,
$|\cos(\phi_0)|<1/\sqrt{\beta R F}$, the crossover time $t_1$ is substantially
larger. In the latter case $t_1$ is the time it takes the particle to
undergo an angular displacement by rotational diffusion, such that the
projected force onto the $x$-axis becomes as large as demanded in the former
case. Only then does the force bring about a lateral displacement that is
compatible or larger than the displacements due to original translational
Brownian motion. Due to this multiplicative coupling of
a diffusive and a ballistic behavior for the angular and the
translational displacements, respectively, the mean square
displacement shows a super-ballistic power-law behavior for $t\gg
t_1$, with  $\langle(x(t)-x_0)^2\rangle\propto t^3$.
 
The intermediate regime is terminated by free rotational Brownian
motion at the second time scale $t_2=D_r^{-1}$, beyond which the
particle motion is diffusive again. As already reported in Ref.~\cite{Cond_Matt},
the long-time translational diffusion constant is given by
\begin{align}
D_L & \equiv\lim_{t\rightarrow\infty}\frac{\langle(x(t)-x_0)^2\rangle}{2t} \notag \\
& =\frac{4}{3} D_r R^2\left[1 + \frac{2}{3}(\beta R F)^2\right]\,.
\label{longtime1}
\end{align}

\begin{figure}
\includegraphics[width=0.45\textwidth]{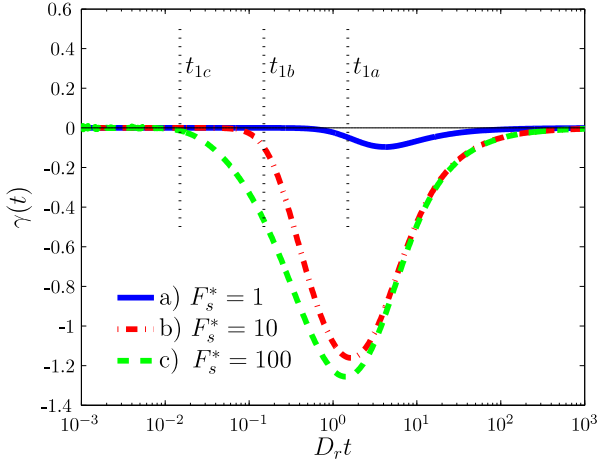}%
\caption{\label{fig:Woelbungmodel11s}(Color online) Kurtosis $\gamma(t)$ of the probability distribution function $\Psi(x,t)$ of a spherical particle with initial orientation angle $\phi_0 = 0.5 \pi$ for the same values of $F_s^* = \beta R F$ as used in Fig.~\ref{fig:Moment2model11s}.  Comparing Figs.~\ref{fig:Moment2model11s} and \ref{fig:Woelbungmodel11s} shows that the time scales $t_{1a}$, $t_{1b}$ and $t_{1c}$ can be extracted from the plots of the kurtosis as well as from the mean square displacement. }
\end{figure}
As can be seen in Fig.~\ref{fig:Woelbungmodel11s}, the beginning of the super-diffusive regime also  shows up in the kurtosis. Here, the deviation from zero clearly indicates the crossover to non-Gaussian behavior. Interestingly, the kurtosis features a pronounced long-time tail. Therefore, the behavior of the particle is still non-Gaussian when its motion  is (nearly) diffusive again. Analysing the analytical result for the kurtosis gives the leading long-time behavior as   
\begin{equation}
\label{tail}
\gamma(t)={\frac{-21{F_s^*}^{4}}{9+12\,{F_s^*}^{2}+4\,{F_s^*}^{4}}}\,(D_\mathrm{r}t)^{-1}+\mathcal{O}\left(\frac{1}{t^2}\right)\,.
\end{equation}
As the amplitude vanishes for $F_s^* =0$, this negative $1/t$ long-time  tail in $\gamma(t)$  proves to be characteristic for self-propelled particles. 

The previous calculation of the displacement moments can also be done for systems in which the self-propelled Brownian particle is exposed to $x$-dependent linear or quadratic potentials. This is illustrated in the following.

\subsection{\label{11slinear}The $(1,1,s)$-model with an additional linear potential}
The Langevin equations in this case are obtained from Eqs.~(\ref{Langevinx1}) and (\ref{Langevinphi1}) by simply inserting a linear potential of the form $U(x)= mgx$ into the first component of  Eq.~(\ref{Langevinx1}). Thus, the motion of a particle that is exposed to gravity is described by the moments   
\begin{equation}
\label{Moment1gravitation}
\langle x(t)-x_0 \rangle =\frac{4}{3}\beta R^2 \left[F \cos(\phi_0)\left(1-e^{-D_rt}\right)-mgD_rt\right]
\end{equation}
and 
\begin{align}
\label{Moment2gravitation}
\left\langle (x(t)-x_0)^2\right\rangle & =  \frac{8}{3}R^2 D_r t + \left(\frac{4}{3} \beta R^2\right)^2 \biggl\{\left(mgD_rt\right)^2    \notag \\ 
& \quad +F^2\biggl[D_rt-1 +e^{-D_rt}+\frac{1}{12} \cos(2\phi_0) \notag \\
& \quad \qquad \times \left(3-4e^{-D_rt}+e^{-4D_rt}\right)\biggr] \notag \\
& \quad  -2Fmg\cos(\phi_0)D_rt\left(1-e^{-D_rt}\right)\biggr\}\,.
\end{align}
While the first moment (Eq.~(\ref{Moment1gravitation})) is a simple superposition of the terms due to the self-propulsion of the particle and the external force, respectively, the mean square displacement (Eq.~(\ref{Moment2gravitation})) has an additional term which depends on both of these forces. 

\begin{figure}
\includegraphics[width=0.45\textwidth]{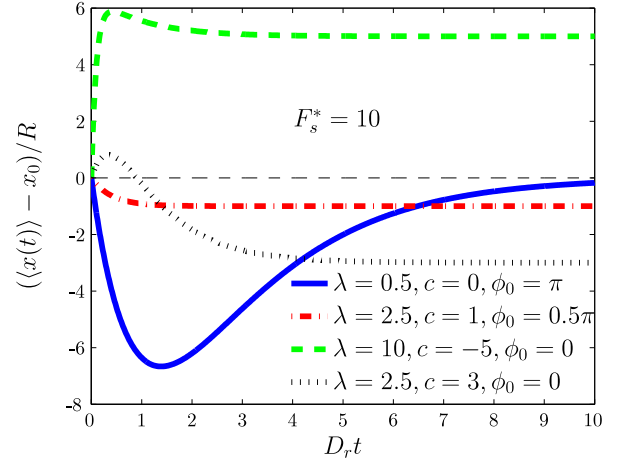}%
\caption{\label{fig:Moment1D1d2oszillator}(Color online) Mean position of a spherical particle with driving force $F_s^* = 10$ that is exposed to an external square potential. The strength of the potential $U(x)=(1/2)kx^2$ is determined by the parameter $\lambda=(4/3)\beta k R^2$. The dimensionless quantity $c=x_0/R$ is the distance between the initial position of the particle and the position of minimal potential given in units of particle radius $R$.}
\end{figure}
\subsection{\label{11ssquare}The $(1,1,s)$-model with an additional square potential}
\begin{figure}
\includegraphics[width=0.45\textwidth]{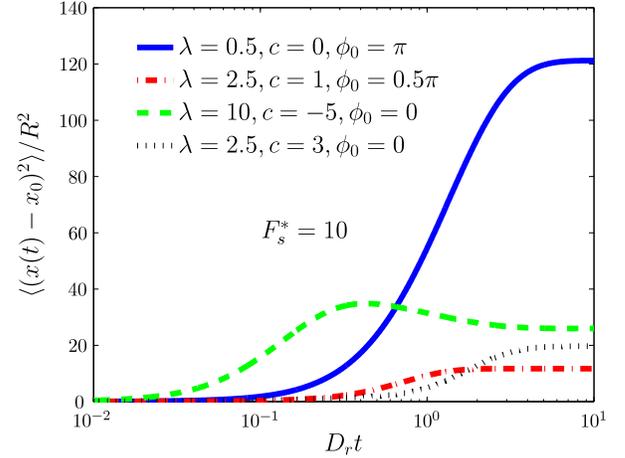}%
\caption{\label{fig:Moment2D1d2oszillator}(Color online) Mean square displacement of a spherical particle that is exposed to an external square potential. The plots refer to the same parameters as the plots in Fig.~\ref{fig:Moment1D1d2oszillator}.}
\end{figure}
Based on the Langevin equations~(\ref{Langevinx1}) and (\ref{Langevinphi1}) one also obtains analytical results for the harmonic oscillator or square potential $U(x)= (1/2) kx^2 $. Using the dimensionless parameter $\lambda  = (4/3) \beta k R^2 $, which determines the strength of the square potential,  the mean position is given by
\begin{equation}
\label{Moment1oszillator}
\langle x(t) \rangle = x_0 e^{-\lambda D_r t}+ \frac{4\beta R^2 F}{3(\lambda -1)} \cos(\phi_0) \left(e^{-D_rt}-e^{-\lambda D_rt}\right)
\end{equation}
and the mean square displacement is calculated as 
\begin{widetext}
\begin{align}
\label{Moment2oszillator}
\langle \left(x(t)-x_0\right)^2  \rangle & = x_0^2\left( e^{-\lambda D_r t}-1\right)^2 + x_0 \frac{8\beta R^2 F}{3(\lambda -1)} \cos(\phi_0) \left(e^{-\lambda D_r t}-1\right) \left(e^{-D_rt}-e^{-\lambda D_r t}\right) + \frac{4 R^2}{3 \lambda}\left(1-e^{-2\lambda  D_rt}\right) \notag \\
& \quad + \left(\frac{4}{3}\beta R^2 F\right)^2 \biggl\{\frac{1}{(\lambda +1) } \left[\frac{1}{2\lambda } \left(1-e^{-2\lambda D_r t}\right)-\frac{1}{(\lambda -1)}\left(e^{-(\lambda +1) D_r t}-e^{-2\lambda D_r t}\right)\right]  \notag \\
& \qquad \qquad \qquad \qquad  +\frac{\cos(2\phi_0)}{(\lambda -3 )} \left[\frac{1}{(2\lambda -4)} \left(e^{-4 D_r t}-e^{-2\lambda D_r t}\right)-\frac{1}{(\lambda -1)}\left(e^{-(\lambda +1) D_r t}-e^{-2\lambda  D_r t}\right)\right]\biggr\} \,.
\end{align}
\end{widetext}
 Figure~\ref{fig:Moment1D1d2oszillator} shows that the mean position of the particle reaches the position of the minimal potential 
  and stays there. Due to the square potential, the mean square displacement (see Fig.~\ref{fig:Moment2D1d2oszillator}) does not diverge as in the cases that have been regarded up to this point. 

\subsection{\label{21s}The $(2,1,s)$-model}
In this section we briefly want to present the results for the $(2,1,s)$-model.  As the one-dimensional case with the theoretically assumed internal force projected onto the $x$-axis was already considered in the preceding subsections, the analytical expressions for the first and second moments based on the two-dimensional Langevin equation~(\ref{Langevinx1}) are  given by superposition of the motion in $x$- and $y$-direction. Thus, using Eqs.~(\ref{Moment11}) and (\ref{Verschiebungsquadrat1}) one obtains the vectorial mean position
\begin{equation}
\label{21smodel1}
\left\langle \mathbf r(t)-\mathbf r_0\right\rangle = \frac{4}{3}\beta FR^2 \left[1-e^{-D_rt}\right]
\begin{pmatrix}
 \cos(\phi_0) \\  \sin(\phi_0)	
\end{pmatrix}
\end{equation}
and the mean square displacement
\begin{align}
\label{21smodel2}
\left\langle (\mathbf r(t)-\mathbf r_0)^2\right\rangle & = \frac{16}{3}R^2 D_r t \notag \\
& \quad + 2\left(\frac{4}{3} \beta FR^2\right)^2 \left[D_rt-1 +e^{-D_rt}\right]\,.
\end{align}
As expected, the $\phi_0$-dependence vanishes in Eq.~(\ref{21smodel2}) due to the free translational motion in the two-dimensional plane. Furthermore, the diffusive term given by the first summand in Eq.~(\ref{21smodel2}) is naturally twice as big as in Eq.~(\ref{Verschiebungsquadrat1}).

\section{\label{einwinkele}Ellipsoidal particle with one orientational degree of freedom}
We now generalize the previous considerations to ellipsoidal particles. To cover the situation depicted in Fig.~\ref{fig:Modelle}(b) we have to take into account that, as opposed to the case of spherical particles, the translational diffusion coefficient is anisotropic, which means that the diffusion tensor
\begin{equation}
\mathbf{D_t}=D_a(\mathbf {\hat{ u}}\otimes \mathbf {\hat{ u}}) +D_b (\mathbf{I} - \mathbf {\hat{ u}}\otimes \mathbf {\hat{ u}}) \\
\label{Diffusionstensor1}
\end{equation} 
has to be applied. Here, $\mathbf{I}$ is the $2\times 2$ unit matrix, $\mathbf {\hat{u}} = (\cos \phi, \sin \phi)$ is the orientation vector, $\otimes$ a dyadic product,  and $D_a$ and  $D_b$, respectively,  indicate the diffusion coefficients  for translation in the direction of the two semi-axes of the ellipsoid. The index $a$ stands for the semi-major axis, while $b$ marks the semi-minor axis. Using the diffusion tensor (Eq.~(\ref{Diffusionstensor1})),  the Langevin equation for the center of mass position of the particle can be written in the form 
\begin{equation}
\frac{\mathrm{d}\mathbf{r}}{\mathrm{d}t} = \beta  \mathbf{D_t} \cdot \left[F \mathbf {\hat{u}} - \mathbf{\nabla} U \right] +  \mathbf{w}\,.
\label{Lanten}
\end{equation}
Due to the anisotropy of the diffusion coefficient we cannot  include the  Gaussian white noise random force exactly in the same way as  in Eq.~(\ref{Langevinx1}). Instead of that, we use the zero mean random noise source $\mathbf{w}(t)$. The variances of the components $i,j \in \{x,y\}$ are given by $\langle w_i(t) w_j(t')\rangle =2 D_t^{ij}(\phi(t))  \delta(t-t')$. Thus, $w_i(t)$ are Gaussian random variables at fixed $\phi(t)$.  
This follows the procedure presented in Ref.~\cite{Han:06} for ``passive'' ellipsoidal particles.

\subsection{\label{11e}The $(1,1,e)$-model}
It can easily be seen from Eq.~(\ref{Diffusionstensor1}) that $\mathbf{D_t} \cdot \mathbf {\hat{u}} = D_a \mathbf {\hat{u}}$. Therefore, in the context of the $(1,1,e)$-model the Langevin equation for the center of mass position $x$ of the self-propelled particle without external potentials can be written as
\begin{equation}
\label{xellipsoid}
\frac{\mathrm{d}x}{\mathrm{d}t} = \beta D_a F \cos(\phi) + w_x\,.
\end{equation}
The Langevin equation for the angle $\phi$,  which is needed in addition to Eq.~(\ref{xellipsoid}), does not differ from Eq.~(\ref{Langevinphi1}). 
For the following calculations, it is convenient to write the diffusion tensor (Eq.~(\ref{Diffusionstensor1})) as  
\begin{equation}
\mathbf{D_t}= \overline{D} \mathbf{I} + \frac{1}{2}\Delta D \begin{pmatrix}
	\cos(2\phi) & \sin (2\phi) \\
	\sin (2\phi) & - \cos(2\phi)
\end{pmatrix}  \,,  
\label{2Diffusionstensor}
\end{equation} 
 where $\overline{D}=1/2(D_a+D_b)$ marks the mean diffusion coefficient and  $\Delta D$ the difference  $D_a - D_b$ between the diffusion coefficients along the long and short axes of the ellipsoid. 

\begin{figure}
\includegraphics[width=0.45\textwidth]{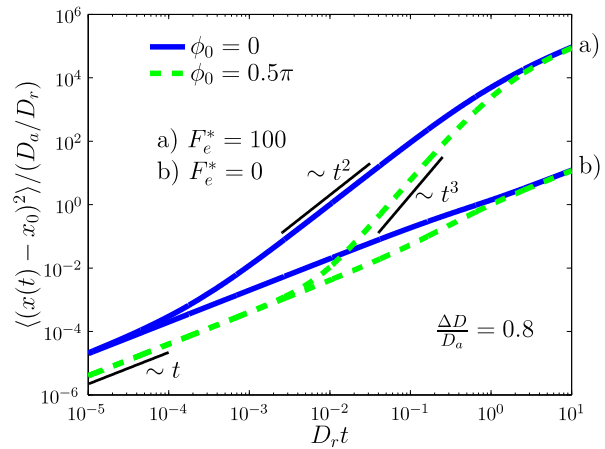}%
\caption{\label{fig:Moment2D1d2ellipsoid}(Color online) Mean square displacement of an ellipsoidal particle with one orientational degree of freedom and anisotropy $\Delta D /D_a = 0.8$. Plots are shown for a) a self-propelled particle with $F_e^*= \beta F_0 \sqrt{D_a/D_r}=100$  and b) a ``passive'' particle with vanishing effective force. The solid lines refer to a parallel and the dashed lines to a perpendicular initial orientation.}
\end{figure}

The mean position of an ellipsoidal particle is given by the first component of Eq.~(\ref{21emodel1}). For the mean square displacement one obtains  the relation
\begin{align}
\label{Verschiebungsquadrate}
\left\langle (x(t)-x_0)^2\right\rangle & = 2\overline{D}t + \frac{\Delta D}{4D_r} \cos(2\phi_0) \left(1-e^{-4D_rt}\right) \notag \\
& \quad + \left(\beta F\frac{D_a}{D_r}\right)^2 \biggl[D_rt-1 +e^{-D_rt}  \notag \\
& \quad +\frac{1}{12} \cos(2\phi_0)\left(3-4e^{-D_rt}+e^{-4D_rt}\right)\biggr]\,.
\end{align}
For an isotropic particle with $\Delta D=0$ Eq.~(\ref{Verschiebungsquadrate}) reduces to the
result for a spherical particle in one dimension (Eq.~(\ref{Verschiebungsquadrat1})), as
expected. For an anisotropic particle the additional (second) term in
Eq.~(\ref{Verschiebungsquadrate}) yields a $\phi_0$-dependence at very early times, in the
regime of bare translational diffusion, which is not present in the
case of an isotropic particle (see the discussion in Sec.~\ref{11s}). This additional term represents the relative orientation of the initial direction of the long axis of the ellipsoidal particle and the direction of the linear channel. The $\phi_0$-dependence can also be seen in Fig.~\ref{fig:Moment2D1d2ellipsoid}, where the strength of the driving force is determined by the parameter $F_e^*= \beta F_0 \sqrt{D_a/D_r}$ for ellipsoidal particles. 

The analytical results for the third and fourth moments, which are needed to calculate skewness and kurtosis,  are more complicated and, therefore, given in the appendix in Eqs.~(\ref{Moment311e}) and (\ref{Moment411e}), respectively.

\subsection{\label{21e}The $(2,1,e)$-model}
Again, we provide the analytical results for the case with two-dimensional translation as well. For an ellipsoidal particle, one obtains the expressions 
\begin{equation}
\label{21emodel1}
\left\langle \mathbf r(t)-\mathbf r_0\right\rangle = \beta F \frac{D_a}{D_r} \left[1-e^{-D_rt}\right]
\begin{pmatrix}
 \cos(\phi_0) \\  \sin(\phi_0)	
\end{pmatrix}
\end{equation}
for the first moment and 
\begin{equation}
\label{21emodel2}
\left\langle (\mathbf r(t)-\mathbf r_0)^2\right\rangle = 4\overline{D}t + 2\left(\beta F \frac{D_a}{D_r}\right)^2 \left[D_rt-1 +e^{-D_rt}\right] 
\end{equation}
for the mean square displacement, respectively. As in the mean square displacement of a spherical particle (Eq.~(\ref{21smodel2})), the $\phi_0$-dependence also vanishes in Eq.~(\ref{21emodel2}).  Furthermore, the  contribution to the diffusive motion due to the initial orientation of the particle  disappears for two-dimensional translation so that the diffusive motion is simply reflected by the term  $4\overline{D}t$ in Eq.~(\ref{21emodel2}).

\section{\label{zweiwinkel} Freely rotating spherical particle}
Up to now, particles with only one orientational degree of freedom have been examined. In this section we transfer our model to particles whose orientation is freely diffusing on the unit sphere. Considering a spherical particle this situation is shown in Fig.~\ref{fig:Modelle}(c).  In the Cartesian lab frame the particle orientation $\mathbf {\hat{ u}} = (\sin\theta \cos\varphi, \sin \theta \sin\varphi, \cos \theta) $ is now given in terms of the two orientation angles $\theta$ and $\varphi$. Using the updated orientation vector the Langevin equation for the center of mass position is identical to Eq.~(\ref{Langevinx1}) if the third component of all vectorial quantities is considered additionally. 

The  orientational probability distribution for the freely diffusing orientation vector~\cite{Dhont_book} is given by 
\begin{equation}
\label{Entwicklungp2}
P(\theta, \varphi, t)  = \sum_{l=0}^\infty \sum_{m=-l}^l e^{-D_rl(l+1)t}\,Y_l^{m *}(\theta_0, \varphi_0) Y_l^m(\theta, \varphi)\,,
\end{equation}
where $Y_l^m$ are the spherical harmonics. 
In Eq.~(\ref{Entwicklungp2}) we use the notation $\theta_0 \equiv \theta(t=0) $ and $\varphi_0 \equiv \varphi(t=0) $ while the star indicates complex conjugation.

\subsection{\label{12s}The $(1,2,s)$-model}
\begin{figure}
\includegraphics[width=0.45\textwidth]{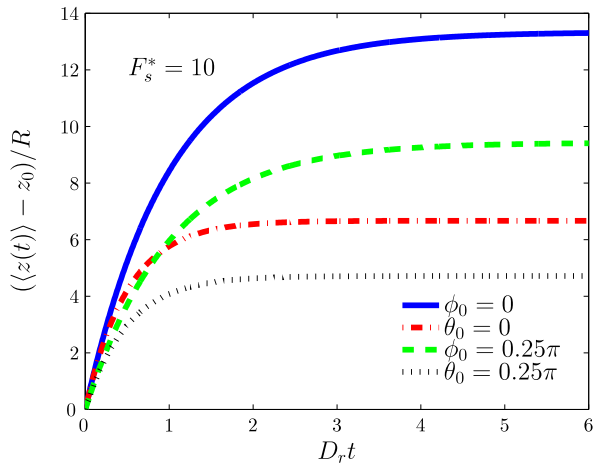}%
\caption{\label{fig:Moment1D1d3}(Color online) Comparison of the mean position of a spherical particle with one and with two orientational degrees of freedom. The graphs for which the initial angle $\phi_0$ is given refer to the $(1,1,s)$-model while the graphs designated by a certain value for the angle $\theta_0$ show the results for the $(1,2,s)$-model. In all cases the effective force is $F_s^*=10$ and the motion in $z$-direction is considered. }
\end{figure}

To eliminate the $\varphi$-dependence in the equation of motion, we choose the $z$-axis to point in the direction of the linear channel, which we consider first. Moreover, we omit external potentials in the following. By calculating $\langle \cos (\theta) \rangle$ via Eq.~(\ref{Entwicklungp2}) the first moment is obtained as 
 \begin{equation}
\label{Moment1z}
\langle z(t) -z_0 \rangle = \frac{2}{3}\beta F R^2 \cos(\theta_0) (1-e^{-2D_rt}) \,.
\end{equation}
The mean position in the $(1,2,s)$-model is very similar to the
same in the $(1,1,s)$-model (see Eq.~\ref{Moment11}). Here, the azimuthal
angle $\theta$ takes the role of the angle $\phi$ in the
$(1,1,s)$-model.  In particular, the two results agree up to
linear order in time $t$, whereas they deviate for longer times due to
the enhanced probability of the sphere with full orientational freedom
to assume a configuration with an orientation pointing in the
direction of the equator.  This, in turn, on average causes a smaller
force component along the $z$-axis and  a smaller plateau-value of the
excursion $\lim_{t\rightarrow\infty}\langle z(t)-z(0)\rangle$, which is illustrated in Fig.~\ref{fig:Moment1D1d3}.

Using Eq.~(\ref{Entwicklungp2}) and, thus, the fact that every function that depends exclusively on  $\theta$ and $\varphi$ can be expanded as a linear combination of spherical harmonics, we also obtain the analytical result for the mean square displacement, which is given by 
\begin{align}
\label{Moment2D1d3}
\langle (z(t)- z_0)^2\rangle & = \frac{8}{3}R^2 D_r t + \left(\frac{2}{9} \beta F R^2\right)^2 \notag \\
& \quad \times \Bigl[  12D_rt-8+9e^{-2D_rt} -e^{-6D_rt} \notag \\
 & \qquad \quad +\cos^2(\theta_0) \left(6-9e^{-2D_rt}+3e^{-6D_rt}\right)\Bigr]\,.
\end{align}

\begin{figure}
\includegraphics[width=0.45\textwidth]{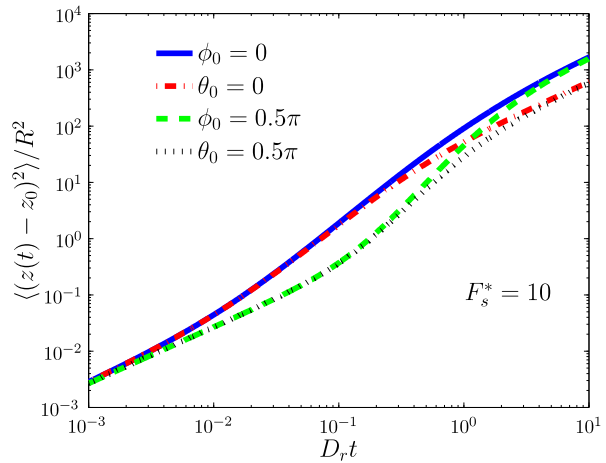}%
\caption{\label{fig:Moment2D1d3}(Color online) Comparison of the mean square displacement of a spherical particle with one and with two orientational degrees of freedom. The remarks in the caption of Fig.~\ref{fig:Moment1D1d3} are also valid for this figure.}
\end{figure}
Comparing Eqs.~(\ref{Moment2D1d3}) and (\ref{Verschiebungsquadrat1}) (as shown in Fig.~\ref{fig:Moment2D1d3}) it turns out that  also the
mean square displacements of the spheres with two or one orientational
degree of freedom in a linear channel are almost identical. In
particular, their functional forms are the same up to second or third
order in time $t$, for the cases of a parallel
($\phi_0=\theta_0\approx 0$) or a perpendicular
($\phi_0=\theta_0\approx\pi/2$) initial configuration, respectively.
Therefore, the crossover time scale from the diffusive to the
super-diffusive regime $t_1$ is exactly the same as in the
$(1,1,s)$-model and given by Eq.~(\ref{ampl-eq}).  The second time scale
from the super-diffusive to the diffusive regime is in the
$(1,2,s)$-model given by $t_2=(2D_r)^{-1}$ and therefore half
as large as in the $(1,1,s)$-model. Concomitantly, the
long-time diffusion constant is smaller as compared to Eq.~(\ref{longtime1}) and
given by
\begin{equation}
D_L=\frac{4}{3} D_r R^2\left[1 + \frac{2}{9}(\beta R F)^2\right]\,.
\end{equation}
As for the long-time limits of the mean positions, the difference to
Eq.~(\ref{longtime1}) reflects the fact that the freely oriented sphere is more
likely oriented perpendicular to the channel axis than the sphere that
is confined to rotate in the plane.

\begin{figure}
\includegraphics[width=0.45\textwidth]{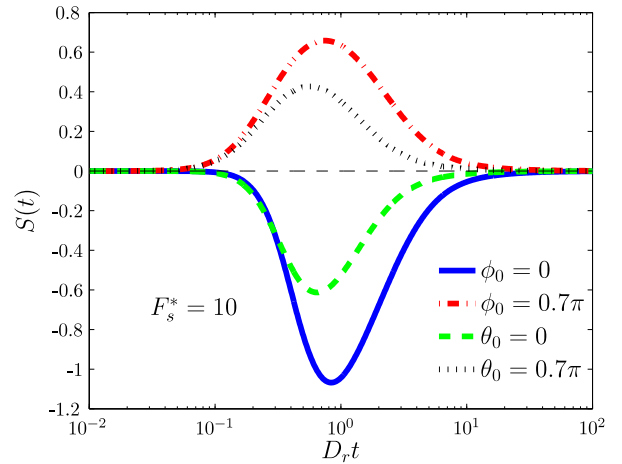}%
\caption{\label{fig:SchiefeD1d3}(Color online) Comparison of the skewness $S(t)$ of the probability distribution function $\Psi(x,t)$ for a spherical particle with one and with two orientational degrees of freedom. The solid and the dash-dotted lines refer to the $(1,1,s)$-model while their counterparts for the $(1,2,s)$-model are given by the dashed and  the dotted lines.}
\end{figure}

\begin{figure}
\includegraphics[width=0.45\textwidth]{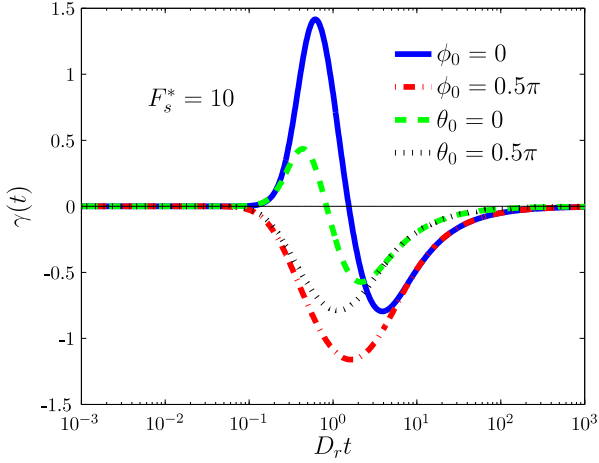}%
\caption{\label{fig:WoelbungD1d3}(Color online) Comparison of the kurtosis $\gamma(t)$ of $\Psi(x,t)$ for a spherical particle with one and with two orientational degrees of freedom. See also the remarks in the caption of Fig.~\ref{fig:SchiefeD1d3}.}
\end{figure}

The non-Gaussian behavior of the self-propelled particle with free orientation on the unit sphere is embodied in the third moment  
 \begin{widetext}
 \begin{align}
\left\langle \frac{(x(t)-x_0)^3}{R^3}\right\rangle & = 16/3\,{ F_s^*}\,\tau\,\cos ( \theta )  ( 1-{e^{-2\,\tau}} )  +{\frac {64}{27}}\,{{ F_s^*}}^{3} \biggl[ -{\frac {41}{96}}\,\cos ( \theta ) +{\frac {1}{96}}\,\cos ( 3\,\theta ) +{\frac {351}{800}}\,\cos ( \theta ) {e^{-2\,\tau}} \notag \\
& \qquad \qquad  -{\frac {3}{160}}\,\cos ( 3\,\theta ) {e^{-2\,\tau}}+1/2\,\cos ( \theta ) \tau -{\frac {1}{800}}\,\cos ( \theta ) {e^{-12\,\tau}}   -{\frac {1}{96}}\,\cos ( \theta ) {e^{-6\,\tau}} \notag \\
& \qquad \qquad  +{\frac {1}{96}}\,\cos ( 3\,\theta ) {e^{-6\,\tau}} +3/10\,\cos ( \theta ) \tau\,{e^{-2\,\tau}}-{\frac {1}{480}}\,{e^{-12\,\tau}}\cos ( 3\,\theta )  \biggr] 
\label{Moment312s}
\end{align}
and in the fourth moment
\begin{align}
\left\langle \frac{(x(t)-x_0)^4}{R^4}\right\rangle & =  {\frac {64}{3}}\,{\tau}^{2}+{\frac {64}{81}}\,{{ F_s^*}}^{2}\tau\, \biggl[ 12\,\tau-8+9\,{e^{-2\,\tau}}-{e^{-6\,\tau}} + ( \cos ( \theta )  ) ^{2} \bigl( 6-9\,{e^{-2\,\tau}}+3\,{e^{-6\,\tau}} \bigr)  \biggr] \notag \\
& \quad +{\frac {2}{893025}}\,{{ F_s^*}}^{4} \biggl[ 27\,{e^{-20\,\tau}}+588\,\cos ( 2\,\theta ) {e^{-12\,\tau}}-16800\,\tau\,\cos ( 2\,\theta ) {e^{-6\,\tau}}-735\,\cos ( 4\,\theta ) {e^{-12\,\tau}} \notag \\
& \qquad  +470400\,{\tau}^{2}+147\,{e^{-12\,\tau}}+1100\,{e^{-6\,\tau}}-5600\,\tau\,{e^{-6\,\tau}}+480298  -481572\,{e^{-2\,\tau}}  \notag  \\
& \qquad -211680\,\tau\,{e^{-2\,\tau}}+105\,{e^{-20\,\tau}}\cos ( 4\,\theta )+60\,{e^{-20\,\tau}}\cos ( 2\,\theta ) +1470\,\cos ( 4\,\theta )\notag  \\
& \qquad -237160\,\cos ( 2\,\theta )  -736960\,\tau-2940\,\cos ( 4\,\theta ) {e^{-2\,\tau}}+249312\,\cos ( 2\,\theta ) {e^{-2\,\tau}}  \notag  \\
& \qquad  +211680\,\tau\,\cos ( 2\,\theta ) {e^{-2\,\tau}}+235200\,\tau\,\cos ( 2\,\theta ) +2100\,\cos ( 4\,\theta ) {e^{-6\,\tau}}-12800\,\cos ( 2\,\theta ) {e^{-6\,\tau}} \biggr] \,.
\label{Moment412s}
\end{align}
\end{widetext}
The  curves for the skewness (see Fig.~\ref{fig:SchiefeD1d3}) and the kurtosis (see Fig.~\ref{fig:WoelbungD1d3}) of the probability distribution function $\Psi(x,t)$ for the $(1,2,s)$-model are obtained by shrinking their counterparts for the $(1,1,s)$-model in $x$-direction as well as in the direction of the $t$-axis. This is very  similar to the findings concerning the mean position of the particle (see Fig.~\ref{fig:Moment1D1d3}).   The extrema of skewness and kurtosis and the  change of sign of the kurtosis, that is observed  for non-perpendicular initial configurations, already occur at smaller times. Obviously, the existence of the negative $1/t$ long-time tail in $\gamma(t)$ (see Sec.~\ref{11s}) is not affected by the number of orientational degrees of freedom of the particle.   

\subsection{\label{22s}The $(2,2,s)$-model}
For more than one translational degree of freedom, if the particle motion takes place in the $xy$-plane, the first and second moments are given by 
\begin{equation}
\label{22smodel1}
\left\langle \mathbf r(t)-\mathbf r_0\right\rangle = \frac{2}{3}\beta F R^2 \left(1-e^{-2D_rt}\right) 
\begin{pmatrix}
 \sin(\theta_0) \cos(\varphi_0) \\   \sin(\theta_0) \sin(\varphi_0)	
\end{pmatrix}
\end{equation}
and
\begin{align}
\label{22smodel2}
\langle (\mathbf{r}(t)- \mathbf{r}_0)^2\rangle & = \frac{16}{3}R^2 D_r t + \left(\frac{2}{9} \beta F R^2\right)^2 \notag \\
& \quad \times \Bigl[  24D_rt-16+18e^{-2D_rt} -2 e^{-6D_rt}\notag \\
 & \qquad \quad  +\sin^2(\theta_0) (6-9e^{-2D_rt}+3e^{-6D_rt})\Bigr]\,,
\end{align}
respectively.
As before, the results for the mean position and the mean square
displacement are almost identical with respect to their
lower-dimensional counterparts. Beside the change from a cosine to a sine in the
last part of Eq.~(\ref{22smodel2}), which is due to the change of accessible
dimensions, the only difference of the mean square
displacements in the $(2,2,s)$- and in the $(1,2,s)$-model consists in an additional factor of $2$ for all terms that do not
depend on $\theta_0$.  

\subsection{\label{32s}The $(3,2,s)$-model}
The $(3,2,s)$-model  is the most general situation concerning a spherical particle. In this case, the mean position of the particle is obtained by adding the $z$-component (Eq.~(\ref{Moment1z})) to Eq.~(\ref{22smodel1}). The mean square displacement is given by 
\begin{align}
\label{32smodel2}
\langle (\mathbf{r}(t)- \mathbf{r}_0)^2\rangle & = 8 R^2 D_r t  \notag \\
& \quad + \frac{1}{2} \left(\frac{4}{3} \beta F R^2\right)^2 \left[2D_rt-1+e^{-2D_rt}\right]\,.
\end{align}
The simplicity of Eq.~(\ref{32smodel2}) can be explained by the fact that no dependence on  the initial orientation can appear due to the completely free motion~\cite{footnote_correlation}.

\section{\label{zweiwinkele}Freely rotating ellipsoidal particle}
To complete our examination of the different model situations, we now consider a freely rotating self-propelled ellipsoidal particle as sketched in Fig.~\ref{fig:Modelle}(d). The $(D,2,e)$-models, where $D \in \{1,2,3\}$ is the number of  translational degrees of freedom, are based on the Langevin equation~(\ref{Lanten}). To transfer this two-dimensional equation to the three-dimensional case regarded here,    the orientation vector $\mathbf {\hat{ u}} = (\sin\theta \cos\varphi, \sin \theta \sin\varphi, \cos \theta) $ is used and the third component is added to all vectorial quantities.  The explicit form of the  diffusion tensor $\mathbf{D_t}$ is given by Eq.~(\ref{Diffusionstensor1}) by means of the updated orientation vector. 

\subsection{\label{12e}The $(1,2,e)$-model}
\begin{figure}
\includegraphics[width=0.45\textwidth]{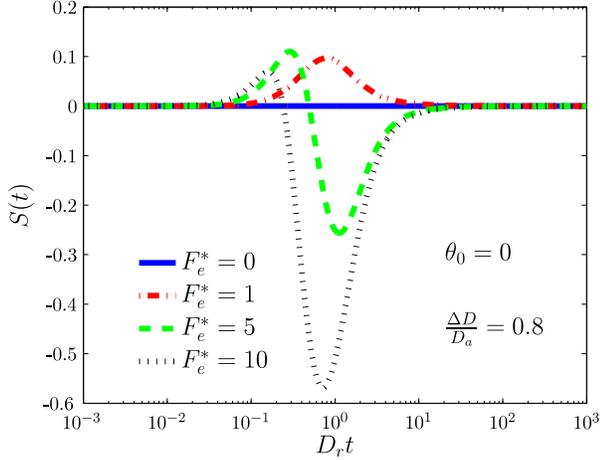}%
\caption{\label{fig:Schiefe12e}(Color online) Skewness $S(t)$ of the probability distribution function $\Psi(x,t)$ for an ellipsoidal self-propelled particle as a function of time. This figure illustrates the analytical results of the $(1,2,e)$-model. While the initial angle $\theta_0$ and the anisotropy $\Delta D/D_a$ are constant as given in the figure, graphs for various values of the effective force $F_e^*$ are shown.  }
\end{figure}

\begin{figure}
\includegraphics[width=0.45\textwidth]{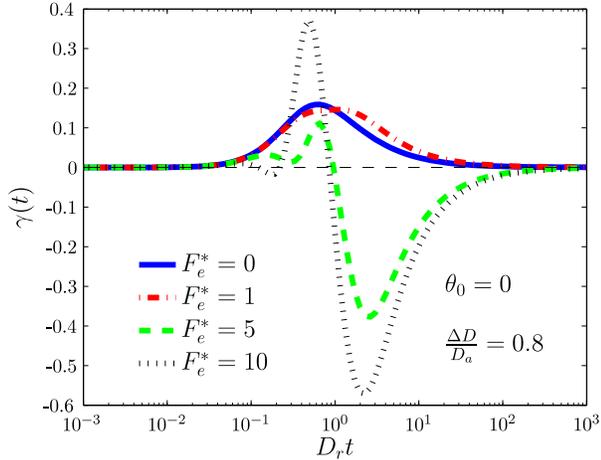}%
\caption{\label{fig:Woelbung12e}(Color online) Kurtosis $\gamma(t)$ of  $\Psi(x,t)$  for an ellipsoidal self-propelled particle as a function of time. The  graphs correspond to the graphs in Fig.~\ref{fig:Schiefe12e} as far as the values of the various parameters are concerned.  }
\end{figure}

As in the previous sections, we begin by considering the case of one-dimensional translational motion without external potentials, i.e., with the $(1,2,e)$-model. Using the ansatz based on spherical harmonics according to Eq.~(\ref{Entwicklungp2}) for an ellipsoidal particle leads to the analytical results
\begin{equation}
\label{12emodel1}
\left\langle z(t)-z_0\right\rangle = \frac{1}{2}\beta F \frac{D_a}{D_r} \left[1-e^{-2D_rt}\right] \cos(\theta_0)
\end{equation}
for the mean position and
\begin{align}
\label{12emodel2}
\langle (z(t)- z_0)^2\rangle  & = 2\overline{D} t + 
    \left(\frac{1}{6} \beta F \frac{D_a}{D_r}\right)^2 \notag \\
& \quad \times  \biggl[  12D_rt-8+9e^{-2D_rt} -e^{-6D_rt} \notag  \\
 & \qquad +\cos^2(\theta_0) \left(6-9e^{-2D_rt}+3e^{-6D_rt}\right)\biggr] \notag \\
& \quad + \frac{\Delta D}{9 D_r} \biggl[-3D_rt-1+e^{-6D_rt} \notag \\
 & \qquad \quad \qquad  + 3 \cos^2(\theta_0) \left( 1-e^{-6D_rt} \right) \biggr] 
\end{align}
for the mean square displacement, respectively. Corresponding expressions for the third and fourth moments and, thus, for skewness and kurtosis were calculated as well. As these are quite lengthy, they are given in the appendix in Eqs.~(\ref{Moment312e}) and (\ref{Moment412e}), respectively, but are presented graphically in Figs.~\ref{fig:Schiefe12e} and \ref{fig:Woelbung12e}. Different regimes of non-Gaussian behavior are manifested by different signs of the kurtosis. For ``passive'' particles with vanishing internal effecitve force $F_e^* =0$ (solid line in Fig.~\ref{fig:Woelbung12e}) no change of sign is observed. The kurtosis is positive and a simple maximum occurs at $t \approx t_2 = (2D_r)^{-1}$. These findings correspond to the results for ``passive'' ellipsoidal particles in two dimensions that were studied in Ref.~\cite{Han:06}. In contrast to simple non-Gaussian behaviour, the situation turns out to be much more complex if self-propelled particles are considered. If the self-propulsion outweighs the effect of the kicks of the solvent particles (dashed line and dotted line in Fig.~\ref{fig:Woelbung12e}), several maxima, minima, and changes of sign induce    a rich structure in $\gamma (t)$ indicating different non-Gaussian behavior at different time scales. The characteristic  $1/t$ long-time  tail observed for spherical particles is also found for ellipsoidal self-propelled  particles.  The skewness $S(t)$ (see Fig.~\ref{fig:Schiefe12e}), which is a measure of the asymmetry of the probability distribution, reveals a higher degree of complexity as well. While $S(t)$ is zero for ``passive'' particles (solid line in Fig.~\ref{fig:Schiefe12e}), this parameter also shows a much richer structure if self-propelled particles are regarded. 

\subsection{\label{22e}The $(2,2,e)$-model}
By simply combining the results for one-dimensional translation in $x$- and in $y$-direction, for the $(2,2,e)$-model we obtain the expressions 
\begin{equation}
\label{22emodel1}
\left\langle \mathbf r(t)-\mathbf r_0\right\rangle = \frac{1}{2}\beta F \frac{D_a}{D_r} \left[1-e^{-2D_rt}\right] \begin{pmatrix}
 \sin(\theta_0) \cos(\varphi_0) \\   \sin(\theta_0) \sin(\varphi_0)	 
 \end{pmatrix}
\end{equation}
for the mean position and 
\begin{align}
\label{22emodel2}
\langle (\mathbf r(t)- \mathbf r_0)^2\rangle  & = 4\overline{D} t + \left(\frac{1}{6} \beta F \frac{D_a}{D_r}\right)^2 \notag \\
& \quad \times \biggl[  24D_rt-16+18e^{-2D_rt} -2e^{-6D_rt}  \notag \\
 &\qquad +\sin^2(\theta_0) \left(6-9e^{-2D_rt}+3e^{-6D_rt}\right)\biggr] \notag \\
 & \quad  + \frac{\Delta D}{9 D_r} \biggl[-6D_rt-2+2e^{-6D_rt} \notag \\
 &\qquad \qquad \quad  + 3 \sin^2(\theta_0) \left( 1-e^{-6D_rt} \right) \biggr]
\end{align}
for the mean square displacement. 

\subsection{\label{32e}The $(3,2,e)$-model}
In a last step, we now add the third translational degree of freedom. Hence, in this subsection we consider the most general case with free translational and rotational motion of an ellipsoidal self-propelled particle. By adding the third component (Eq.~(\ref{12emodel1})) to Eq.~(\ref{22emodel1}) we obtain the result for the mean position. 
The analytical expression for the mean square displacement is simply given by 
\begin{align}
\label{32emodel2}
\langle (\mathbf r(t)- \mathbf r_0)^2\rangle  &  = 6\overline{D} t - \Delta D t \notag \\
& \quad + \frac{1}{2} \left(\beta F \frac{D_a}{D_r}\right)^2 \left[  2D_rt-1+e^{-2D_rt}\right]\,.
\end{align}
To explain the simplicity of this result we want to point to the short discussion after Eq.~(\ref{32smodel2}).

\section{\label{Schluss}Conclusion}
In conclusion, we have analytically solved the Brownian dynamics of an 
anisotropic self-propelled particle
in different geometries by presenting explicit results for the first four 
moments of the probability distribution function for displacements. The particle is driven along an axis which itself  
fluctuates according to
rotational Brownian dynamics. After a transient regime which is 
characterized by two distinct time scales,
there is  diffusive behavior for long times. The results for the long-time diffusion constants $D_L$ for the different groups of model situations are given in Table~\ref{tab:DL}.   
 \begin{table}
 \caption{\label{tab:DL}Long-time diffusion constant $D_L$ for the different model situations. Using the definition in  Eq.~(\ref{longtime1}) the analytical results are directly obtained from the respective results for the mean square displacement.
  }
 \begin{ruledtabular}
 \begin{tabular}{lc}
 Model & Long-time diffusion constant \\
 \hline
 $(D,1,s)$  &   $(4/3)D_r R^2\left[1 + (2/3)(\beta F R)^2\right]$  \\
 $(D,1,e)$  & $ \overline{D} + 1/(2 D_r)\left(\beta F \left(\overline{D} + (1/2)\Delta D \right)\right)^2 $ \\
 $(D,2,s)$  & $(4/3) D_r R^2\left[1 + (2/9)(\beta F R)^2\right] $ \\
 $(D,2,e)$  & $ \overline{D} - (1/6)\Delta D + 1/(6 D_r)\left(\beta F \left(\overline{D} + (1/2)\Delta D \right)\right)^2$ 
 \end{tabular}
 \end{ruledtabular}
 \end{table}
 For intermediate
times,  non-Gaussian behavior is revealed by a non-vanishing kurtosis 
in the particle displacement which decays as $1/t$ for long times $t$.
For special initial conditions (nearly perpendicular initial orientation and large effective forces), we find a superdiffusive transient regime
where the mean square displacement scales with an exponent 3 in time. 
The analytical results can be used to compare with experimental systems of, 
e.g.,  swimming bacteria
or self-propelled colloidal particles. Any deviations point to
the importance of hydrodynamic interactions with the substrate and with 
neighbouring
particles at finite density.

It would be interesting to generalize the analysis towards various 
situations.
First of all, hydrodynamic interactions were neglected in our studies. 
While this is justified in the bulk,
hydrodynamic interactions become important at finite densities 
\cite{Yeomans} and may significantly
influence the distribution of the mean square displacements 
\cite{Goldstein_PRL_2009}. 
If the particle is moving close to a substrate and the substrate is 
hydrodynamically not flat,
then hydrodynamic interactions play a significant role, too~\cite{Berke:08}. Second, it 
would be interesting to include
  an additional torque in the Langevin equations of motion. This leads 
to circular motion in two
dimensions~\cite{Teeffelen_PRE} while for three spatial dimensions helical motion is expected as also suggested by a slightly different model presented in Ref.~\cite{Friedrich:09}.  
 Next, a non-Gaussian noise~\cite{Strefler:09} in the Langevin equations might 
be relevant for modeling
real swimming objects \cite{Friedrich,Friedrich:09}. Furthermore,
the actual propulsion mechanism was modelled just by an effective force. A consideration of more details 
about the actual propulsion mechanism
might be necessary to analyse short-time dynamics in more depth. 
In addition to that, the model migth be transferred to more complicated geometries~\cite{Lindner:08} such as ratchets~\cite{Tailleur:09,Galajda:07,Reichhardt:08}, for example. 
Finally, the collective dynamics~\cite{Munk,Franosch,Romanczuk:09} of many swimmers will 
lead to further effects like swarming,
swirling and jamming. While at high densities hydrodynamic interactions 
are expected to play a minor role, 
the direct particle-particle interactions get relevant and should be 
incorporated
in theory \cite{Wensink:08} and simulation \cite{Baer}.

\begin{acknowledgments}
We thank T. Lubensky, H. H. Wensink, J. K. G. Dhont, J. Dunkel, and J. Yeomans 
for 
helpful discussions.
This work was supported by the DFG (SFB TR6 - C3).
\end{acknowledgments}

\begin{widetext}
\appendix*
\section{Further analytical results}
In the following, we summarize the analytical results for the third and fourth moments of an ellipsoidal particle. For reasons of clarity, we use the notation $F_e^*= \beta F_0 \sqrt{D_a/D_r}$, the ratio $\delta = \Delta D /D_a$ and the scaled time $\tau = D_r t$. For the $(1,1,e)$-model the third moment is given by 
\begin{align}
\left\langle \frac{(x(t)-x_0)^3}{(D_a/D_r)^{3/2}}\right\rangle & = {{ F_e^*}}^{3} \biggl[ -{\frac {45}{8}}\,\cos ( \phi ) +\frac {1}{24}\,\cos ( 3\,\phi ) +{\frac {17}{3}}\,\cos ( \phi ) {e^{-\tau}}-\frac {1}{16}\,\cos ( 3\,\phi ) {e^{-\tau}}+3\,\cos ( \phi ) \tau-\frac {1}{24}\,\cos ( \phi ) {e^{-4\,\tau}}\notag \\
& \qquad \qquad+\frac {1}{40}\,\cos ( 3\,\phi ) {e^{-4\,\tau}}+\frac {5}{2}\,\cos ( \phi ) \tau\,{e^{-\tau}}-{\frac {1}{240}}\,{e^{-9\,\tau}}\cos ( 3\,\phi )  \biggr] \notag \\
& \quad+6\,{ F_e^*}\, \left( 1-\frac {1}{2}\,{ \delta} \right)  \biggl[ \cos ( \phi ) \tau-\cos ( \phi ) \tau\,{e^{-\tau}} \biggr] \notag \\
& \quad+3\,{ F_e^*}\,{ \delta}\, \biggl[ \frac {5}{8}\,\cos ( \phi ) +{\frac {5}{72}}\,\cos ( 3\,\phi ) -\frac {2}{3}\,\cos ( \phi ) {e^{-\tau}}-\frac {1}{16}\,\cos ( 3\,\phi ) {e^{-\tau}} \notag \\
& \qquad \qquad \qquad +{\frac {13}{720}}\,{e^{-9\,\tau}}\cos ( 3\,\phi ) -\frac {1}{2}\,\cos ( \phi ) \tau\,{e^{-\tau}}+\frac {1}{24}\,\cos ( \phi ) {e^{-4\,\tau}}-\frac {1}{40}\,\cos ( 3\,\phi ) {e^{-4\,\tau}} \biggr] 
\label{Moment311e}
\end{align} 
and the fourth moment is 
\begin{align}
\left\langle \frac{(x(t)-x_0)^4}{(D_a/D_r)^2}\right\rangle & =  {\frac {1}{100800}}\,{{ F_e^*}}^{4} \biggl[ 15\,{e^{-16\,\tau}}\cos ( 4\,\phi ) -504000\,\tau\,{e^{-\tau}}-1134000\,\tau +1644300+168\,\cos ( 2\,\phi ) {e^{-9\,\tau}} \notag \\
& \qquad \qquad \qquad \quad -319200\,\cos ( 2\,\phi ) +420\,\cos ( 4\,\phi ) {e^{-4\,\tau}} +525\,\cos ( 4\,\phi ) +2100\,{e^{-4\,\tau}}-1646400\,{e^{-\tau}} \notag \\
& \qquad \qquad \qquad \quad  -1568\,\cos ( 2\,\phi ) {e^{-4\,\tau}}-840\,\cos ( 4\,\phi ) {e^{-\tau}}+151200\,\tau\,\cos ( 2\,\phi ) -3360\,\tau\,\cos ( 2\,\phi ) {e^{-4\,\tau}} \notag \\
& \qquad \qquad \qquad \quad   +320600\,\cos ( 2\,\phi ) {e^{-\tau}}-120\,\cos ( 4\,\phi ) {e^{-9\,\tau}} +168000\,\tau\,\cos ( 2\,\phi ) {e^{-\tau}}+302400\,{\tau}^{2} \biggr] \notag \\
& \quad+{{ F_e^*}}^{2} \left( 1-1/2\,{ \delta} \right)  \biggl[ 3\,\tau\,\cos ( 2\,\phi ) -12\,\tau-4\,\tau\,\cos ( 2\,\phi ) {e^{-\tau}}+12\,\tau\,{e^{-\tau}}+\tau\,\cos ( 2\,\phi ) {e^{-4\,\tau}}+12\,{\tau}^{2} \biggr] \notag \\
& \quad+6\,{{ F_e^*}}^{2}{ \delta}\, \biggl[ {\frac {6917529027641081855}{9223372036854775808}}\,\tau +{\frac {76861433640456465}{4611686018427387904}}\,\tau\,\cos ( 2\,\phi ) {e^{-4\,\tau}} -1/6\,\tau\,\cos ( 2\,\phi ) {e^{-\tau}}   \notag \\
& \qquad \qquad \quad  +{\frac {9223372036854775807}{6917529027641081856}}\,{e^{-\tau}}+{\frac {461168601842738791}{332041393326771929088}}\,\cos ( 4\,\phi ) {e^{-4\,\tau}}  \notag \\
& \qquad \qquad \quad-{\frac {23058430092136939525}{166020696663385964544}}\,\cos ( 2\,\phi ) -{\frac {2305843009213693951}{110680464442257309696}}\,{e^{-4\,\tau}}   \notag \\
& \qquad \qquad \quad -{\frac {39199331156632797169}{46485795065748070072320}}\,{e^{-16\,\tau}}\cos ( 4\,\phi )  -{\frac {374699488997225267}{103762935414616227840}}\,\cos ( 2\,\phi ) {e^{-9\,\tau}}    \notag  \\
& \qquad \qquad \quad   -{\frac {2767011611056432759}{830103483316929822720}}\,\cos ( 2\,\phi ) {e^{-4\,\tau}}+1/4\,\tau\,\cos ( 2\,\phi )   -{\frac {11}{720}}\,\cos ( 4\,\phi ) {e^{-\tau}}  \notag \\
& \qquad \qquad \quad -{\frac {48422703193487572987}{36893488147419103232}}+{\frac {374699488997225267}{145268109580462718976}}\,\cos ( 4\,\phi ) {e^{-9\,\tau}}  \notag  \\
& \qquad \qquad \quad+{\frac {3026418949592973313}{20752587082923245568}}\,\cos ( 2\,\phi ) {e^{-\tau}}+1/2\,\tau\,{e^{-\tau}}+{\frac {16140901064495857663}{1328165573307087716352}}\,\cos ( 4\,\phi )  \biggr] \notag \\
& \quad+12\, \left( 1-1/2\,{ \delta} \right) ^{2}{\tau}^{2}+3\,{ \delta}\, \left( 1-1/2\,{ \delta} \right) \cos ( 2\,\phi ) \tau\, \left( 1-{e^{-4\,\tau}} \right) \notag \\
& \quad +3/4\,{{ \delta}}^{2} \left[ \tau-1/4+1/4\,{e^{-4\,\tau}}+1/48\,\cos ( 4\,\phi )  ( 3-4\,{e^{-4\,\tau}}+{e^{-16\,\tau}} )  \right] \,.
\label{Moment411e}
\end{align}
The corresponding results for a freely rotating ellipsoidal particle ($(1,2,e)$-model) are the third moment
\begin{align}
\left\langle \frac{(x(t)-x_0)^3}{(D_a/D_r)^{3/2}}\right\rangle & = {\frac {1}{2400}}\,{{ F_e^*}}^{3} \biggl[ -1025\,\cos ( \theta ) +25\,\cos ( 3\,\theta ) +1053\,\cos ( \theta ) {e^{-2\,\tau}}  -45\,\cos ( 3\,\theta ) {e^{-2\,\tau}}+1200\,\cos ( \theta ) \tau \notag \\
& \qquad \qquad  \qquad -3\cos ( \theta ) {e^{-12\tau}}  -25\cos ( \theta ) {e^{-6\tau}}+25\cos ( 3\theta ) {e^{-6\tau}}  +720\cos ( \theta ) \tau{e^{-2\tau}}-5{e^{-12\tau}}\cos ( 3\theta )  \biggr]  \notag \\
& \quad +6\,{ F_e^*}\, \left( 1-1/2\,{ \delta} \right)  \biggl[ 1/2\,\cos ( \theta ) \tau-1/2\,\cos ( \theta ) \tau\,{e^{-2\,\tau}} \biggr] \notag \\
& \quad +3{ F_e^*}{ \delta} \biggl[ {\frac {7}{36}}\cos ( \theta ) +\frac {1}{36}\cos ( 3\theta )   +{\frac {1}{72}}\cos ( \theta ) {e^{-6\tau}}-{\frac {1}{72}}\cos ( 3\theta ) {e^{-6\tau}}  -\frac {1}{6}\cos ( \theta ) \tau  -{\frac {43}{200}}\cos ( \theta ) {e^{-2\tau}}  \notag \\
& \qquad \qquad  \qquad +{\frac {1}{150}}\,\cos ( \theta ) {e^{-12\,\tau}}+{\frac {1}{90}}\,{e^{-12\,\tau}}\cos ( 3\,\theta )  -\frac {1}{40}\,\cos ( 3\,\theta ) {e^{-2\,\tau}}-\frac {1}{10}\,\cos ( \theta ) \tau\,{e^{-2\,\tau}} \biggr] 
\label{Moment312e}
\end{align}
and the fourth moment
\begin{align}
 \left\langle \frac{(x(t)-x_0)^4}{(D_a/D_r)^2}\right\rangle &  =  {\tau}^{2}  \biggl[ 1/3\,{{ F_e^*}}^{4}+4\,{{ F_e^*}}^{2} ( 1-1/2\,{ \delta} ) -4\,{ \delta}\, ( 1-1/2\,{ \delta} ) \notag \\
& \quad \qquad  -{\frac {295147905179352817853}{442721857769029238784}}\,{{ F_e^*}}^{2}{ \delta}   +12\, ( 1-1/2\,{ \delta} ) ^{2} +1/3\,{{ \delta}}^{2} \biggr]  \notag \\
& \quad +  \tau  \biggl[ 6\,{{ F_e^*}}^{2}{ \delta}\, \biggl( {\frac {16971004547812787156293}{79689934398425262981120}}   +{\frac {184467440737095508157}{55782954078897684086784}}\,{e^{-6\,\tau}} \notag \\
& \quad \qquad \qquad +{\frac {184467440737095508157}{18594318026299228028928}}\,\cos ( 2\,\theta ) {e^{-6\,\tau}}   +{\frac {230584300921369395}{9223372036854775808}}\,{e^{-2\,\tau}} \notag \\
& \quad \qquad \qquad -{\frac {230584300921369395}{9223372036854775808}}\,\cos ( 2\,\theta ) {e^{-2\,\tau}}  +{\frac {147573952589676420931}{5312662293228350865408}}\,\cos ( 2\,\theta )  \biggr)   \notag \\
& \quad \qquad +3\,{{ \delta}}^{2} \biggl( -1/9\,\cos ( 2\,\theta )  +{\frac {11}{135}}-{\frac {1}{189}}\,{e^{-6\,\tau}}   -{\frac {1}{63}}\,\cos ( 2\,\theta ) {e^{-6\,\tau}} \biggr) \notag \\
& \quad \qquad+{{ F_e^*}}^{4} \biggl( {\frac {3}{20}}\,\cos ( 2\,\theta ) {e^{-2\,\tau}}+1/6\,\cos ( 2\,\theta ) -{\frac {3}{20}}\,{e^{-2\,\tau}} -{\frac {1}{252}}\,{e^{-6\,\tau}}   -{\frac {1}{84}}\,\cos ( 2\,\theta ) {e^{-6\,\tau}}-{\frac {47}{90}} \biggr) \notag \\
& \quad \qquad +1/6\,{{ F_e^*}}^{2} ( 1-1/2\,{ \delta} )  \biggl( -10+6\,\cos ( 2\,\theta ) +9\,{e^{-2\,\tau}}-9\,\cos ( 2\,\theta ) {e^{-2\,\tau}}   +3\,\cos ( 2\,\theta ) {e^{-6\,\tau}}+{e^{-6\,\tau}} \biggr) \notag \\
& \quad \qquad +2/3\,{ \delta}\, ( 1-1/2\,{ \delta} )  \biggl( 3\,\cos ( 2\,\theta )  +1-3\,\cos ( 2\,\theta ) {e^{-6\,\tau}}-{e^{-6\,\tau}} \biggr)  \biggr]   \notag \\
& \quad +{{ F_e^*}}^{4} \biggl[ {\frac {1}{2400}}\,\cos ( 2\,\theta ) {e^{-12\,\tau}} +{\frac {1}{9600}}\,{e^{-12\,\tau}}+{\frac {1}{672}}\,\cos ( 4\,\theta ) {e^{-6\,\tau}}-{\frac {4}{441}}\,\cos ( 2\,\theta ) {e^{-6\,\tau}}  \notag \\
& \quad \qquad  -{\frac {121}{720}}\,\cos ( 2\,\theta ) -{\frac {1}{480}}\,\cos ( 4\,\theta ) {e^{-2\,\tau}}+{\frac {53}{300}}\,\cos ( 2\,\theta ) {e^{-2\,\tau}}+{\frac {1}{960}}\,\cos ( 4\,\theta ) \notag \\
& \quad \qquad  +{\frac {11}{14112}}\,{e^{-6\,\tau}}+{\frac {4901}{14400}}   +{\frac {1411740617885935}{73786976294838206464}}\,{e^{-20\,\tau}}-{\frac {273}{800}}\,{e^{-2\,\tau}} \notag \\
& \quad \qquad  +{\frac {1}{23520}}\,{e^{-20\,\tau}}\cos ( 2\,\theta ) -{\frac {1}{1920}}\,\cos ( 4\,\theta ) {e^{-12\,\tau}}   +{\frac {1}{13440}}\,{e^{-20\,\tau}}\cos ( 4\,\theta )  \biggr] \notag \\
& \quad   +6\,{{ F_e^*}}^{2}{ \delta}\, \biggl[ {\frac {304002342334733405771}{2361183241434822606848}}\,{e^{-2\,\tau}}   +{\frac {1475739525896763653647}{318759737593701051924480}}\,\cos ( 2\,\theta )  \notag \\
& \quad \qquad \qquad  +{\frac {26235369349275805845}{18889465931478580854784}}\,\cos ( 4\,\theta ) {e^{-12\,\tau}}   -{\frac {5247073869855161169}{18889465931478580854784}}\,{e^{-12\,\tau}}  \notag \\
& \quad \qquad \qquad  +{\frac {491913175298921373631}{141670994486089356410880}}\,\cos ( 4\,\theta )    +{\frac {24595658764946071489}{49584848070131274743808}}\,\cos ( 4\,\theta ) {e^{-6\,\tau}}  \notag \\
& \quad \qquad \qquad  +{\frac {29809938423114633542725}{3123845428418270308859904}}\,\cos ( 2\,\theta ) {e^{-6\,\tau}}   -{\frac {807229520665529965278287}{6375194751874021038489600}}  \notag \\
& \quad \qquad \qquad  -{\frac {271536072765004556323}{21250649172913403461632}}\,\cos ( 2\,\theta ) {e^{-2\,\tau}}   -{\frac {807045053224792883}{166020696663385964544}}\,\cos ( 4\,\theta ) {e^{-2\,\tau}}  \notag \\
& \quad \qquad \qquad  -{\frac {2951479051793528185637}{23139595766061261547110400}}\,{e^{-20\,\tau}}   -{\frac {2951479051793528185637}{10412818094727567696199680}}\,{e^{-20\,\tau}}\cos ( 2\,\theta )  \notag \\
& \quad \qquad \qquad  -{\frac {5247073869855161169}{4722366482869645213696}}\,\cos ( 2\,\theta ) {e^{-12\,\tau}}   -{\frac {2951479051793528185637}{5950181768415752969256960}}\,{e^{-20\,\tau}}\cos ( 4\,\theta )  \notag \\
& \quad \qquad \qquad   -{\frac {673306158690398658161}{390480678552283788607488}}\,{e^{-6\,\tau}} \biggr]  \notag \\
& \quad  +3\,{{ \delta}}^{2} \biggl[ {\frac {1}{280}}\,{e^{-20\,\tau}}\cos ( 4\,\theta ) +{\frac {1}{490}}\,{e^{-20\,\tau}}\cos ( 2\,\theta )  -{\frac {1}{84}}\,\cos ( 4\,\theta ) {e^{-6\,\tau}}-{\frac {19}{1800}}-{\frac {37}{1323}}\,\cos ( 2\,\theta ) {e^{-6\,\tau}}  \notag \\
& \qquad \qquad  +{\frac {1}{120}}\,\cos ( 4\,\theta ) +{\frac {7}{270}}\,\cos ( 2\,\theta ) +{\frac {17}{1764}}\,{e^{-6\,\tau}}+{\frac {9}{9800}}\,{e^{-20\,\tau}} \biggr] \,.
\label{Moment412e}
\end{align} 
\end{widetext}
\bibliography{journals,literature}

\begin{thebibliography}{68}
\bibitem [{foo(){\natexlab{a}}}]{footnote_forcefree}%
  \BibitemOpen
  \bibinfo {note} {It is important to note that the motion of swimmers is
  force-free. Hence the force entering into the Langevin equation is an
  effective one which describes the propulsion mechanism on
  average.}\BibitemShut {Stop}%
\bibitem [{foo(){\natexlab{b}}}]{footnote_Erdmann}%
  \BibitemOpen
  \bibinfo {note} {Alternatively, self-propelled particles have been modeled by
  damped Brownian dynamics with a Rayleigh friction term~\cite{Erdmann:02} or
  by prescribing a velocity field on the surface of the particle due to surface
  deformations~\cite{Downton:09}.}\BibitemShut {Stop}%
\bibitem [{foo(){\natexlab{c}}}]{footnote_correlation}%
  \BibitemOpen
  \bibinfo {note} {We note that, due to the independence of the initial
  configuration, Eq.~(\ref{32smodel2}) can also be derived directly from the
  correlation function $\langle \hat{ u}(t) \cdot \hat{ u}_0 \rangle =
  \exp{(-2D_rt)}$~\citep{Doi_Edwards_book}.}\BibitemShut {Stop}%
\end{thebibliography}

\end{document}